\documentclass{cernrep} 
\usepackage{texnames}
\usepackage[T1]{fontenc}
 \def\bc{\begin{center}}          \def\ec{\end{center}}
\usepackage{subfig}
\usepackage{graphicx}
\usepackage{rotating}
\usepackage{authblk}

\def\lambdabar{\protect\@lambdabar}
\def\@lambdabar{%
\relax \bgroup
\def\@tempa{\hbox{\raise.73\ht0
\hbox to0pt{\kern.2\wd0\vrule width.7\wd0
height.1pt depth.1pt\hss}\box0}}%
\mathchoice{\setbox0\hbox{$\displaystyle\lambda$}\@tempa}%
{\setbox0\hbox{$\textstyle\lambda$}\@tempa}%
{\setbox0\hbox{$\scriptstyle\lambda$}\@tempa}%
{\setbox0\hbox{$\scriptscriptstyle\lambda$}\@tempa}%
\egroup }

\title{Summary of strong-field QED Workshop}

\author[1]{M. Altarelli}
\author[2]{R. Assmann}
\author[2]{F. Burkart}
\author[2]{B. Heinemann\footnote{Also at University of Freiburg}}
\author[3]{T. Heinzl}
\author[4]{T. Koffas}
\author[5]{A.R. Maier}
\author[6]{D.~Reis}
\author[2]{A. Ringwald}
\author[7]{M. Wing\footnote{Also at DESY, Hamburg. E-mail: m.wing@ucl.ac.uk}}
\affil[1]{Max Planck Institute for the Structure and Dynamics of Matter, Hamburg, Germany}
\affil[2]{Deutsches Elektronen-Sychrotron, Hamburg, Germany}
\affil[3]{University of Plymouth, Plymouth, UK}
\affil[4]{Carleton University, Ottawa, Canada}
\affil[5]{University of Hamburg, Hamburg, Germany}
\affil[6]{SLAC National Accelerator Laboratory, Menlo Park, USA}
\affil[7]{University College London, London, UK}

\vspace{2cm}

\begin{document}

\maketitle

\begin{abstract}{
A workshop, ``Probing strong-field QED in electron--photon interactions'', was held in DESY, Hamburg in August 2018, gathering 
together experts from around the world in this area of physics as well as the accelerator, laser and detector technology that 
underpins any planned experiment. The aim of the workshop was to bring together experts and those interested in measuring QED 
in the presence of strong fields at and above the Schwinger critical field.  The pioneering experiment, E144 at SLAC, measured 
multi-photon absorption in Compton scattering and $e^+e^-$ pair production in electron--photon interactions but never reached the 
Schwinger critical field value. With the advances in laser technology, in particular, new experiments are being considered which 
should be able to measure non-perturbative QED and its transition from the perturbative regime. This workshop reviewed the 
physics case and current theoretical predictions for QED and even effects beyond the Standard Model in the interaction of a 
high-intensity electron bunch with the strong field of the photons from a high-intensity laser bunch. The world's various electron 
beam facilities were reviewed, along with the challenges of producing and delivering laser beams to the interaction region. 
Possible facilities and sites that could host such experiments were presented, with a view to experimentally realising the Schwinger 
critical field in the lab during the 2020s.
}
\end{abstract}

\vspace{5cm}
\centerline{\Large \bf April 2019}

\tableofcontents
\pagebreak[4]
\




\section{Introduction}
B. Heinemann, M. Wing.

A workshop, "Probing strong-field QED in electron--photon interactions", was held in DESY in August 2018 to bring together experts and 
those interested in measuring QED in the presence of strong fields at and above the Schwinger critical field.  The pioneering experiment, 
E144 at SLAC, measured multi-photon absorption in Compton scattering and $e^+e^-$ pair production in electron--photon interactions 
but never reached the Schwinger critical field value. With the advances in laser technology, in particular, new experiments are being 
considered which should be able to measure non-perturbative QED and its transition from the perturbative regime.  This write-up gives a 
summary of the workshop presentations and discussions, with a brief outline, below, of the various experiments.

A summary of previous current and planned experiments to measure strong-field QED using high-power laser pulses and high-energy 
electron bunches is shown in Table~\ref{tab:exps}.  In the late 1990s, the E144 collaboration were the first experiment to try and 
measure the effects of strong QED 
fields using the high energy electrons ($46.6$\,GeV) from the Final Focus Test Beam at SLAC, but were limited by the laser power 
and stability at that time.  The only other experiment which has recently taken data is at the Astra Gemini laser at the 
Rutherford Appleton Laboratory.  Here a high-power laser pulse is used to create electron bunches through laser-driven wakefield 
acceleration (LWFA) as well as for the $e\gamma$ collision.  This allows precision control of the timing of the laser pulse and electron 
bunch, but currently suffers from low statistics and an electron beam that is not highly reproducible.  Future experiments at Astra Gemini 
to assess these issues are planned.

Experiments planned over the next $\sim 5$\,years are also listed in Table~\ref{tab:exps}.  As at the Astra Gemini laser, experiments 
planned using the Extreme Light Infrastructure (ELI), will rely on a high-power laser for the the production of electron bunches using LWFA 
as well as their collision.  The highest power lasers will be available at ELI ($10$\,PW) and electrons of up to $10$\,GeV are expected.  
The other two planned experiments, LUXE and using FACET-II, rely on high-quality RF electron beams in the $10-20$\,GeV range 
at European XFEL and SLAC, respectively, which will be brought into collision with a dedicated high-power laser system.  Two stages 
of experimentation are foreseen for LUXE in which the laser power is upgraded.

\pagestyle{empty}
\begin{sidewaystable}
\caption{ \it Summary of previous, current and planned experiments.}
\label{tab:exps}
\begin{center}
\begin{tabular}{c|ccc|ccc|c} \hline\hline
\rule{0pt}{1.0em}%
Name & Year & $a_0^{\rm max}$ & $\chi^{\rm max}$  & \multicolumn{3}{|c|}{Laser} & $e$ beam\\[2pt] 
 & & & & Power [TW] & A [$\mu$m$^2$] & $\lambda$ [nm] & \\\hline\hline
\rule{0pt}{1.0em}%
E144 & 1990s & 0.36 & 0.3 & 1 & 100 & $500-1000$ &$E_e=46.6$\,GeV, RF \\
Astra Gemini & 2017 & 9.0 & $0.15-0.3$ & 200 & $25$ &$800-1000$ & $E_e\approx 1-2$\,GeV  LWFA\\
ELI-NP & 2018 & 100 & 10 & 10000  & $10$ & $800$ & up to 10\,GeV, LWFA \\ 
FACET-II & 2020 & 7.2 & 0.85 & 20 & $10$  & $800$ & $E_e=10$\,GeV, RF \\
LUXE-I &  2021 &1.5&  0.3 &  10  &$100$& $800$ &$E_e=17.5$\,GeV, RF\\
LUXE-II & 2025 &6.8 &  1.4&  200 & $100$  &$800$ &$E_e=17.5$\,GeV, RF\\
\hline\hline
\end{tabular}
\end{center}
\end{sidewaystable}
\pagestyle{plain}

In the following, the session convenors have written an overall summary, briefly describing the contributions, with each session organised 
into the following sections.
The theoretical background and recent advances presented at the workshop is described in Section~\ref{sec:theory}.  The section covers 
the latest QED calculations and methods as well as possible signatures for physics beyond the Standard Model in $e\gamma$ collisions.   
Sections~\ref{sec:acc} and~\ref{sec:lasers} review the status of the various electron beams and high-power lasers around the world.  In 
Section~\ref{sec:det} the challenges facing detectors and some ideas on the solutions are discussed.  During the workshop, a session was 
dedicated to other experiments which are strongly synergetic with investigations of QED in $e\gamma$ collisions; this is reviewed in 
Section~\ref{sec:syn}.  Finally a brief summary and conclusions is given in Section~\ref{sec:summary}.


\section{Theory}
\label{sec:theory}
T. Heinzl, A. Ringwald

\subsection{Introduction}

Strong field QED (SFQED) originates from standard QED when it makes physical sense to decompose the photon field $A^\mu$ into a dominant classical background $\bar{A}^\mu$ and a `small' quantum fluctuation $a^\mu$. There are four distinct types of electromagnetic backgrounds that can be classified invariantly and in suitable frames become purely electric, magnetic, electric-magnetic or null. The latter case typically refers to plane waves or their long-wavelength limit (constant crossed fields), both of which are used to model intense laser fields (neglecting transverse focussing). Due to the technological progress in laser science leading to unprecedented magnitudes in power, intensity and field strength, plane wave backgrounds have become relevant for a theoretical description of QED in the presence of intense laser fields (sometimes referred to as high-intensity QED). It has to be emphasised that the background field \emph{approximation} just described assumes a large (macroscopic) number of photons comprising the background and that this number basically does not change (absence of backreaction). Once these assumptions become invalid, one has to modify the theory and use full QED for \emph{all} electromagnetic modes. 

QED is a relativistic quantum field theory, hence characterised by the universal constants $c$ and $\hbar$ (usually set to unity) and the parameters $e$ and $m$ representing electron charge and mass (excluding further leptons for simplicity). From these four fundamental constants one can form the typical QED electric field,
\begin{equation}
   E_S = m^2 c^3 / e\hbar \simeq 1.3 \times 10^{18} \; \mbox{V/m} \; ,
\end{equation}
called the  Sauter--Schwinger field. In this field an electron gains energy $mc^2$ across a Compton wavelength, $\lambdabar_e = \hbar/mc$. As is well known since the days of Klein's paradox, localizing an electron to within a Compton wavelength implies pair creation as first explained by Sauter \cite{Sauter:1931zz}, Heisenberg, Euler \cite{Heisenberg:1935qt} and elegantly worked out by Schwinger \cite{Schwinger:1951nm}. The challenge is to realise a field of magnitude comparable to $E_S$ across a \emph{macroscopic} distance, $d  \gg \lambdabar_e$. If one can do so and, say, create a constant electric field $E$ (in the lab frame), then, according to Schwinger \cite{Schwinger:1951nm}, the pair creation rate per unit volume will be of the form
\begin{equation} \label{SCHWINGER}
  \mathcal{R} \sim \alpha\, m^2 E^2 \exp (-\pi E_S/E) = \alpha\, m^2 E^2 \exp(-\pi m^2 /eE) \; .
\end{equation}
Three remarks are in order: First, for the time being, one has to live with $E \ll E_S$, thus exponential suppression of the rate (but see below). Second, the exponent has the charge $e$ in the denominator so that the answer (\ref{SCHWINGER}) cannot be obtained in perturbation theory, i.e.\ by the usual expansion in powers of the QED coupling $\alpha \equiv e^2/4\pi = 1/137$. Schwinger has thus provided a \emph{nonperturbative} result, of which there are few. Third, a purely electric field represents a particular invariance class called `hyperbolic' which has substantial `pair creativity' once large enough. Purely magnetic and null (plane wave) fields, however, have zero pair creativity no matter how strong in magnitude they are. This may be seen by considering the generalisation of (\ref{SCHWINGER}) to arbitrary frames \cite{Nikishov:1969tt}, where one has
\begin{equation} \label{NIKISHOV}
  \mathcal{R} \sim \mathcal{E}\mathcal{B} \, \coth(\pi \mathcal{B}/\mathcal{E}) 
  \, \exp(-\pi m^2 /e\mathcal{E}) \; ,
\end{equation}
in terms of the invariant field magnitudes
\begin{equation}
  \mathcal{E} := \left( \sqrt{\mathcal{S}^2 + \mathcal{P}^2} - 
  \mathcal{S}\right)^{1/2} 
  \; , \quad
 \mathcal{B} := \left( \sqrt{\mathcal{S}^2 + \mathcal{P}^2} +
  \mathcal{S}\right)^{1/2}  \; ,
\end{equation}
first introduced by Heisenberg and Euler \cite{Heisenberg:1935qt}. These in turn are expressed via the standard scalar and pseudo-scalar (Schwinger) invariants,
\begin{equation} \label{SP}
  \mathcal{S} := - \frac{1}{4} F_{\mu\nu} F^{\mu\nu} = (E^2 - B^2)/2 
  \; , \quad
  \mathcal{P} := - \frac{1}{4} F_{\mu\nu} \tilde{F}^{\mu\nu} = 
  \mathbf{E} \cdot \mathbf{B} \; ,
\end{equation}
with $\mathbf{E}$ and $\mathbf{B}$ the ambient electric and magnetic field in the lab. 

The Nikishov rate (\ref{NIKISHOV}) coincides with Schwinger's when $\mathcal{B} = 0$ and vanishes in the purely magnetic case ($\mathcal{E} = 0$) and for null fields, $\mathcal{E} = \mathcal{B} = 0$. Interestingly, though, when $\mathcal{E} \ne 0$, a magnetic field provides an enhancement factor of the form $\mathcal{B} \coth(\pi \mathcal{B}/\mathcal{E})$. Producing pairs in fields with $\mathcal{E} = 0$ (or $\mathcal{E} \ll \mathcal{B}$) requires a `stimulus', typically in the form of a \emph{probe particle}, say an electron. The Schwinger--Nikishov exponent will then be modified through the replacement of the invariant $\mathcal{E}$ by the electric field  $\mathcal{E}'$ experienced by the probe in its instantaneous rest frame, 
\begin{equation}
  \mathcal{E} \to \mathcal{E}' := 
  (p_\mu F^{\mu\alpha} F_{\alpha\nu} p^\nu)^{1/2}/m \; ,
\end{equation} 
where $p = (E_p, \mathbf{p})$ denotes the probe momentum. Typically there will be corrections to this leading order exponent which characterise the details of the probe and background field (frequencies, intensity and finite size effects, etc.). For probe electrons it is customary to introduce the inverse Schwinger--Nikishov exponent or `quantum nonlinearity parameter',
\begin{equation}
 \chi := \mathcal{E}'/E_S \; .
\end{equation}
Let us assume that the background electromagnetic field is characterised by a dominating momentum $k$. For an intense plane wave modelling a laser, $k$ is a light-like wave vector with $k^2 = \omega^2 - \mathbf{k}^2 = 0$. We then introduce the dimensionless energy ratio 
\begin{equation}
  b_0 := \frac{k\cdot p}{m^2} \simeq \frac{2\gamma \omega}{m} \; ,
\end{equation}
with $2 \gamma \omega$ being the laser frequency `seen' by an electron probe in a head-on collision at high energy ($\gamma = E_p/m \gg 1$). Forming the ratio of invariants $\chi$ and $b_0$ leads to the `classical intensity parameter',
\begin{equation}
  a_0 := \frac{\chi}{b_0} = \frac{eE}{m \omega} \; ,
\end{equation}
where the last expression has been evaluated in the lab frame and represents the energy gain of an electron across a laser wavelength, $\lambdabar = 1/\omega$, in units of the electron rest energy, $m$. [Note that different (sub)communities have different names for this parameter such as $x$, $\xi$, $\eta$, etc.] Hence, electrons in a field of magnitude $E$ become relativistic when $a_0 \simeq 1$. This corresponds to a laser intensity of roughly $10^{18}$\,W/cm$^2$. 

The general pair production probability may be written as 
\begin{equation} \label{GEN.RATE}
  \mathcal{R} \sim \exp \{- f(a_0, b_0)/\chi \} \; ,
\end{equation}
with a process dependent function $f$. As an example we quote the asymptotic rate for Breit--Wheeler (BW) pair production \cite{Breit:1934zz} (when $a_0 \gg 1$, $\chi \ll 1$) as recently given by Hartin et al.\, \cite{Hartin:2018sha} who find  
\begin{equation} \label{RITUS.CORNER}
  f(a_0) = -\frac{8}{3} \left\{ 1 - \frac{1}{15 a_0^2} + O(a_0^{-4}) \right\} \; .
\end{equation} 

\subsection{Trident pair production and sub-processes}

\begin{figure}[ht!]
\centering
\includegraphics[scale=0.25]{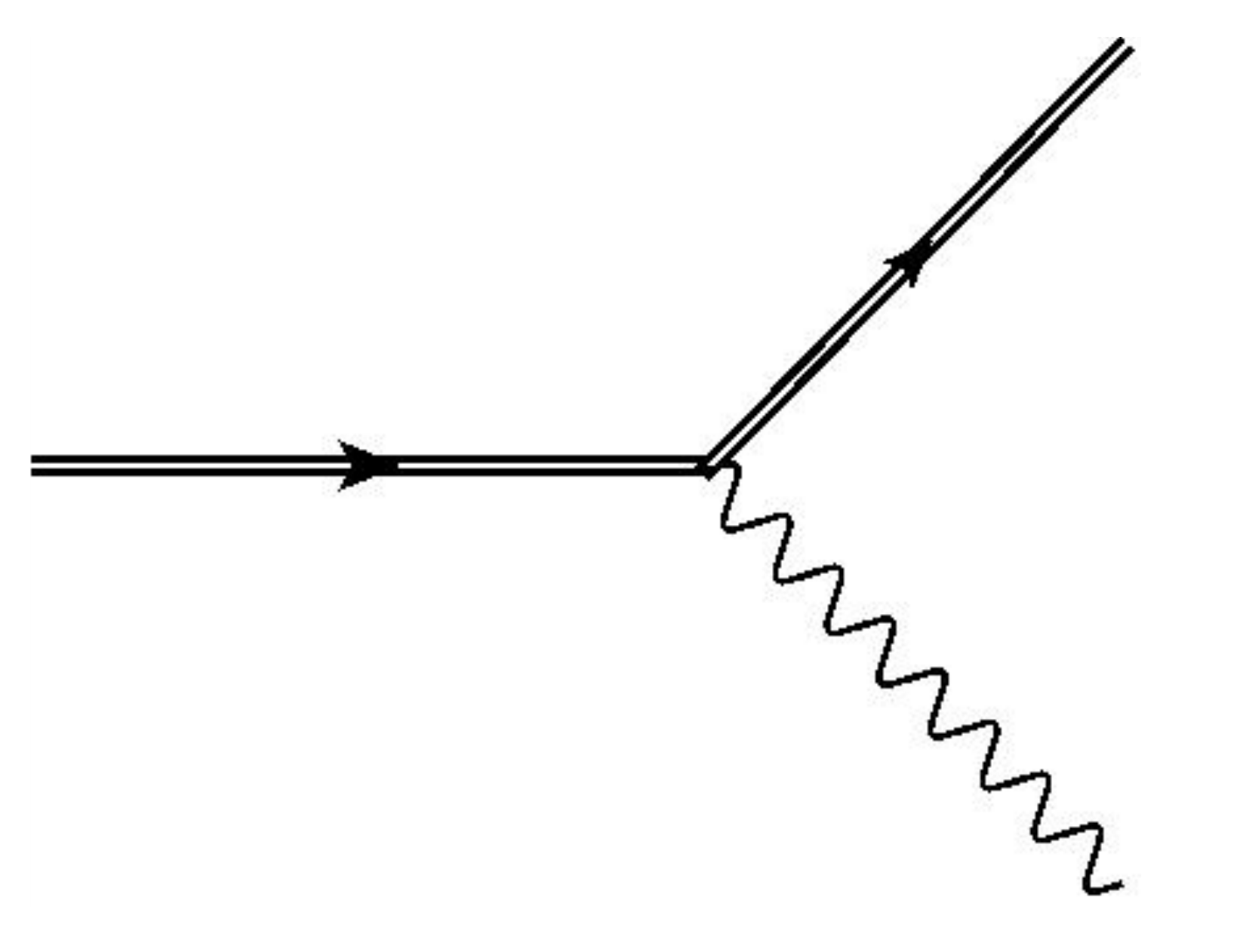} \hspace{2cm}
\includegraphics[scale=0.25]{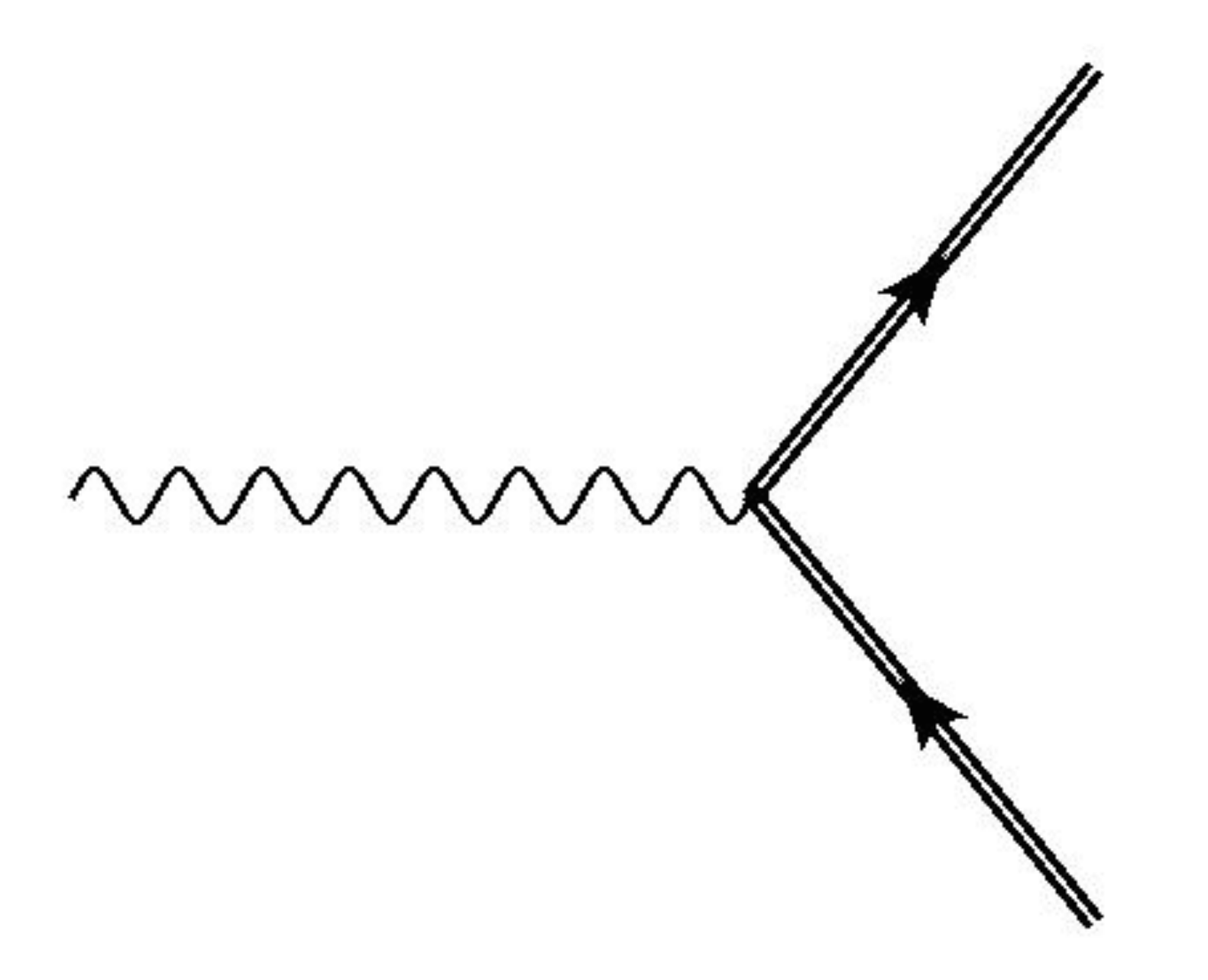}
\caption{Left: Nonlinear Compton scattering (photon emission). Right: Nonlinear Breit--Wheeler pair production. Double lines denote laser dressed (Volkov) elecrons.} 
\label{fig:NLC.BW} 
\end{figure}

As discussed in the introduction, the SLAC experiment E-144~\cite{prd:60:092004} has for the first (and only) time measured two strong-field QED processes, namely nonlinear Compton (NLC) scattering and multi-photon BW pair production. These are shown in Fig.~\ref{fig:NLC.BW}, the double lines denoting Volkov electrons, i.e.\ electrons dressed through the interaction with the laser photons. 

\begin{figure}[ht]
\centering
\includegraphics[scale=0.25]{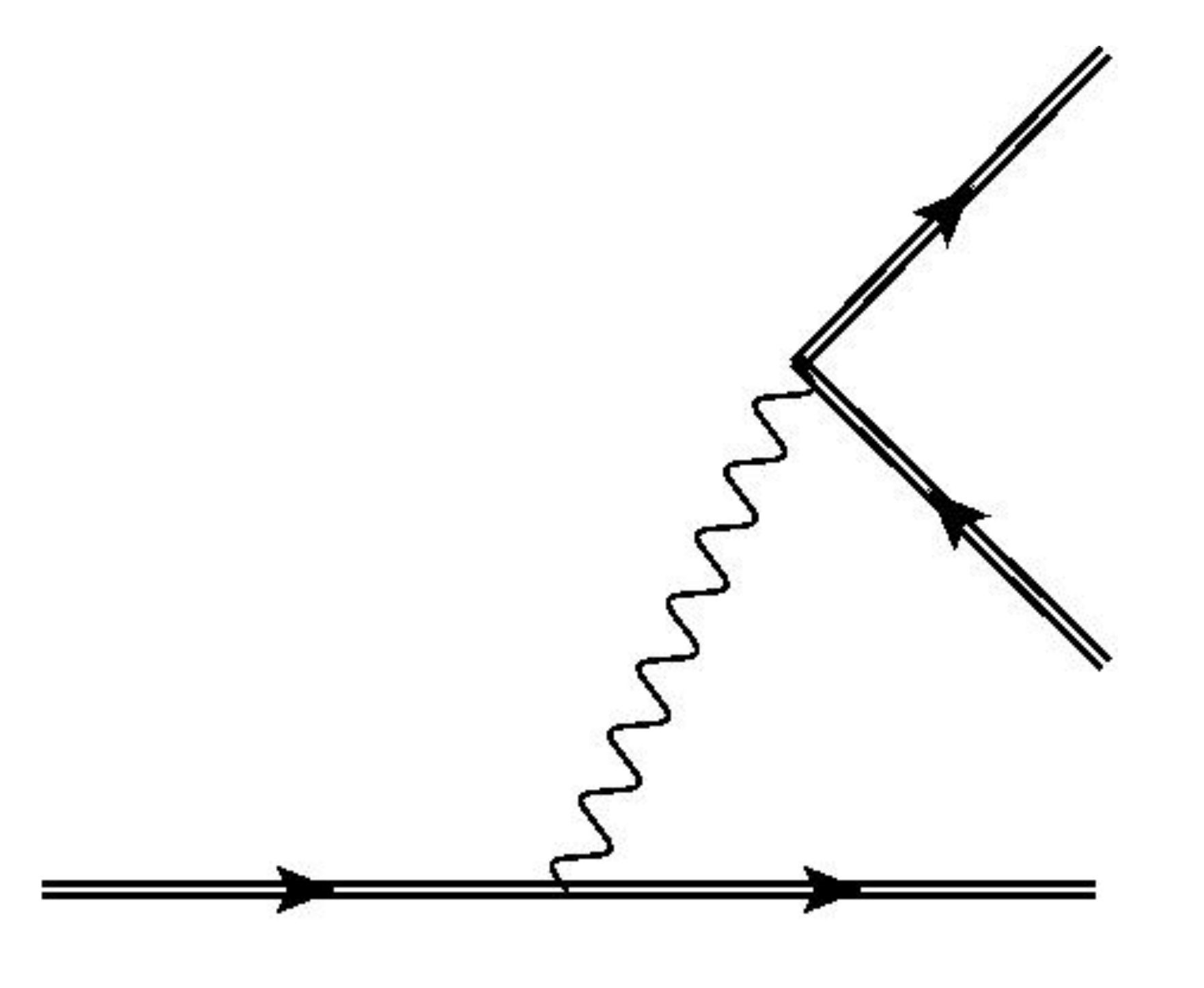} 
\caption{The trident process for Volkov electrons. Cutting through the internal photon line produces the processes of Fig.~\ref{fig:NLC.BW}.} 
\label{fig:trident} 
\end{figure}

E-144 has observed NLC scattering, that is the emission of a photon resulting from electron laser interactions with up to $n_L=4$ laser photons \cite{Bula:1996st} through processes of the form $e + n_L \gamma_L \to e' + \gamma$. These were generated experimentally by bringing 46.6\,GeV electrons into collision with a laser with intensity parameter $a_0 \simeq 0.4$. In a second step, the backscattered laser photons of energy 29.2\,GeV produced electron positron pairs via multi-photon BW processes, $\gamma + n_L \gamma_L \to e^+ e^-$. Getting above threshold required the participation of $n_L=5$ laser photons which was experimentally confirmed \cite{Burke:1997ew}. In principle, this two-step process (NLC followed by BW), can also proceed in a single step through the exchange of a \emph{virtual} photon via the (laser dressed) trident process, its naming stemming from the fact that an initial electron ends up in a three-particle final state (an outgoing electron and a pair, see Fig.~\ref{fig:trident}). Using the Weizs\"acker-Williams approximation, E-144 estimated the direct trident pair production rate to be suppressed by three orders of magnitude compared to the two-step process measured (see Fig.~5 in \cite{Burke:1997ew}). 

\begin{figure}[ht]
\centering
\includegraphics[scale=0.4]{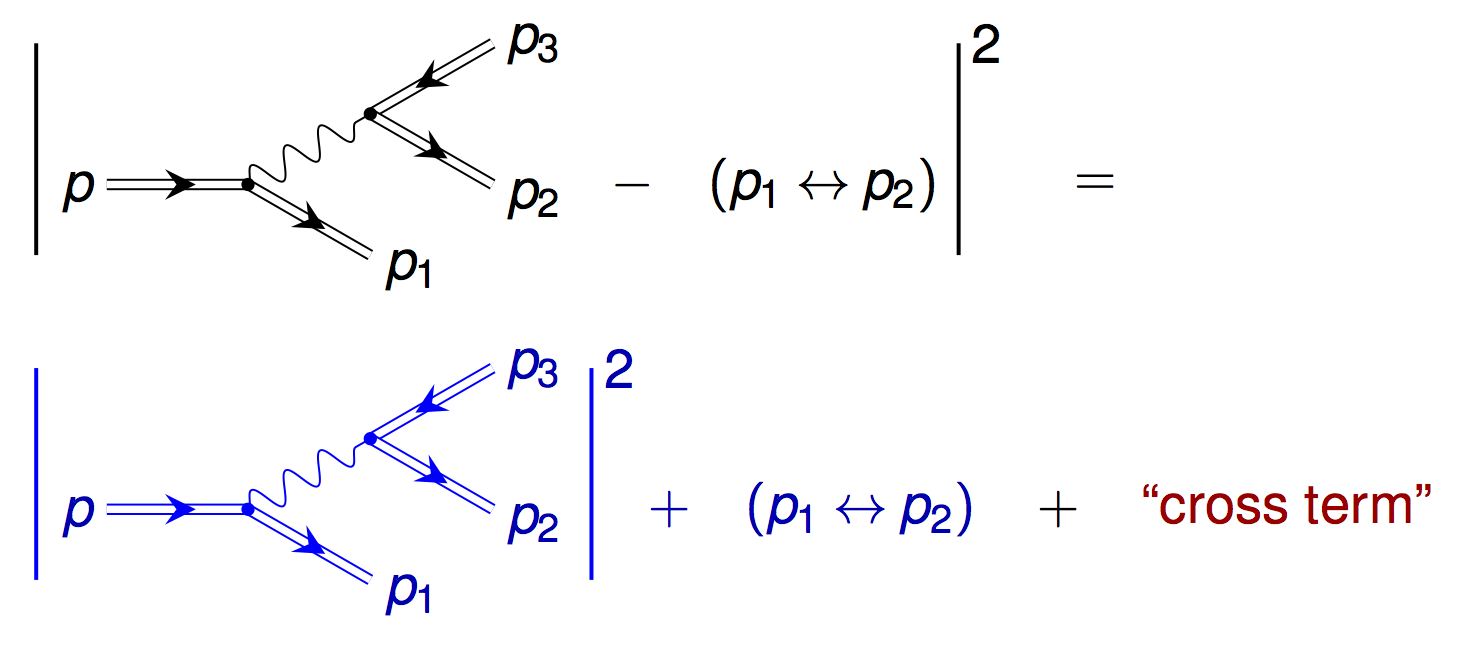} 
\caption{Decomposition of the trident rate into direct and exchange parts (in blue and red, respectively.)} 
\label{fig:tridentRate} 
\end{figure}

The last decade or so has seen substantial progress in the analytical treatment of the trident process and was summarised in three contributions during the workshop by TORGRIMSSON, K\"AMPFER and MACKENROTH. The trident process is notorious for the technical challenges it presents. These have essentially two reasons: First, the final state has three particles, thus an intricate phase-space, in particular if all final momenta (or even spins) are accounted for. Second, the virtual (internal) photon can go on-shell and become real as there is phase-space available due to energy--momentum exchange with the background. Equivalently, the processes of Fig.~\ref{fig:NLC.BW} are possible sub-processes of laser-dressed trident while they cannot happen in the absence of a (laser) background. As a result, the total trident pair creation rate decomposes into a one-step and a two-step part (the latter often referred to as the cascade process), $\mathcal{R} = \mathcal{R}_1 + \mathcal{R}_2$. Switching off the background ($a_0 = 0$) implies $\mathcal{R}_2 = 0$, so one expects the virtual channel, $\mathcal{R}_1$, to dominate in the regime where the background is perturbative, 
\begin{equation}
  \mathcal{R}_1 \gg \mathcal{R}_2 \; , \quad a_0 \ll 1 \; .
\end{equation} 
For strong backgrounds, $a_0 \gg 1$, the situation changes, and the two-step (cascade) channel becomes dominant \cite{Gonoskov:2014mda}. Thus, there is a transitional regime, $a_0 = O(1)$, where one expects one-step and two-step processes to be of comparable magnitude. Covering basically the whole range of parameter space, Greger TORGRIMSSON (FSU and HI Jena) reported on a recent in-depth investigation employing both analytical and numerical tools \cite{Dinu:2017uoj}. Before presenting results, 
he disentangled one-step and two-step processes and addressed a further complication: because there are two identical particles (electrons) in the final state there is also an exchange contribution as illustrated in Fig.~\ref{fig:tridentRate}. 

The full decomposition of the trident rate thus may be written as 
\begin{equation}
  \mathcal{R} = \mathcal{R}_1 + \mathcal{R}_2 = \mathcal{R}_1^\mathrm{dir} + 
  \mathcal{R}_1^\mathrm{exc} + \mathcal{R}_2^\mathrm{dir} \; , 
\end{equation}
where $\mathcal{R}_1^\mathrm{exc}$ denotes the one-step exchange contribution and so on. It is this particular term that has been calculated for the first time in \cite{Dinu:2017uoj}. In previous works it had been neglected.  

In the transition regime, $a_0 \sim 1$, $\chi \ll 1$, and for a finite sinusoidal wave train containing $N$ cycles, one finds $\mathcal{R}_2 \sim N \mathcal{R}_1^\mathrm{dir} \sim N \mathcal{R}_1^\mathrm{exc}$. Thus, for long pulses, the two-step (cascade) process dominates by a factor of $N$ while sub-dominant one-step direct and exchange contributions are of equal magnitude. This suggests that one-step becomes important and non-negligible for short, few-cycle pulses. Interestingly, using saddle point methods, one can find an analytic result for the total rate, which is of the form (\ref{GEN.RATE}) with
\begin{equation}
  f(a_0) = 4 a_0 \left\{ (2 + a_0^2) \, \sinh^{-1} (1/a_0) - \sqrt{1 + a_0^2} \right\}
\end{equation}
Fitting the rate to the form $\exp (-C/\chi)$, SLAC E-144 finds $C_\mathrm{SLAC} = 2.4 \pm 0.5$ which compares favourably to the speaker's answer $C_\mathrm{DT} \simeq 2.46$ \cite{Dinu:2017uoj}. However, as E-144 was at the borderline between the perturbative and non-perturbative regimes, one cannot unambiguously distinguish between exponential and power-law behaviour of the form $\mathcal{R} \sim a_0^{8/b_0}$. 

For general values of $a_0$ and $\chi$ one has to rely on numerical results which show that $\mathcal{R}_1^\mathrm{dir} \sim \mathcal{R}_1^\mathrm{exc}$ for $\chi$ values between 10 and 30, so that the exchange term can safely be neglected for $\chi \gtrsim 10^2$ only.     

\begin{figure}[ht]
\centering
\includegraphics[scale=0.9]{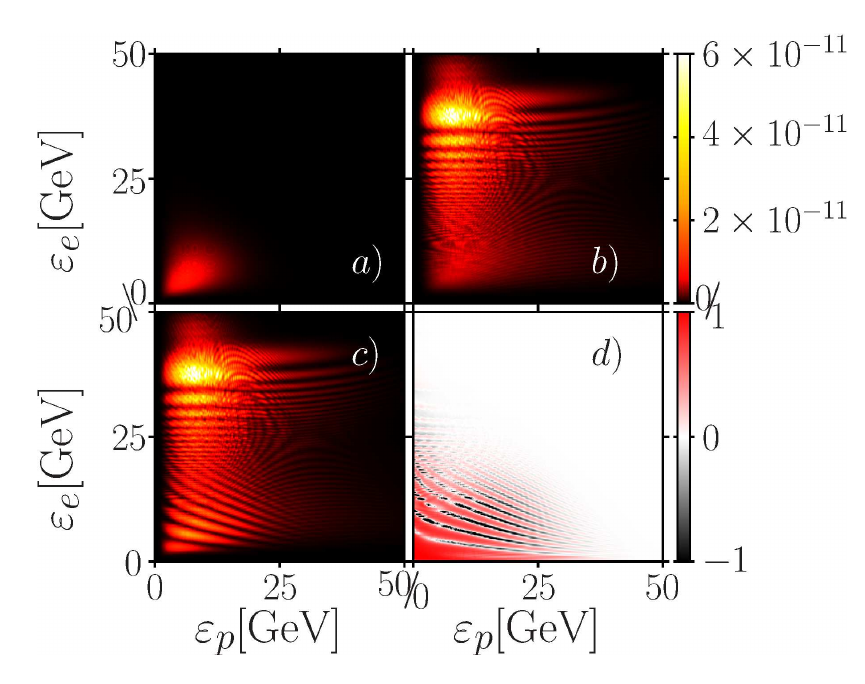} 
\caption{Differential rates for trident (sub)processes: (a) $\mathcal{R}_1$ for one-step, (b) $\mathcal{R}_2$ for two-step (cascade) and (c) total rate $\mathcal{R} = \mathcal{R}_1 + \mathcal{R}_2$ as a function of electron and positron energies, $\epsilon_e$ and $\epsilon_p$, respectively. Scenario:  Collision of a 100 GeV electron with a laser pulse ($a_0 = 11$, $\chi = 13$). The relative error, $\mathcal{R}_1/\mathcal{R}$, of the cascade approximation is shown in (d).} 
\label{fig:Mackenroth} 
\end{figure}

The results presented by Felix MACKENROTH (MPIPKS, Dresden) based on \cite{Mackenroth:2018smh} seem nicely consistent with the above. In particular, they confirm that in general the one-step contribution can be neglected ($\mathcal{R}_1 \ll \mathcal{R}_2$) unless $a_0 \sim \chi \sim O(10)$, which corresponds to high electron beam energies, see Fig.~\ref{fig:Mackenroth}. Thus, in this parameter regime, the one-step trident contribution needs to be determined and included. This is also consistent with older results of the beamstrahlung community which state that in the `deep quantum regime', $\chi \gg 1$, the virtual channel will become dominant \cite{Chen:1989ss}. It was also reassuring to see that the number of positrons observed at SLAC E-144 is consistent with the reported findings on the trident process when E-144 parameters are adopted. 

\begin{figure}[ht]
\centering
\includegraphics[scale=0.5]{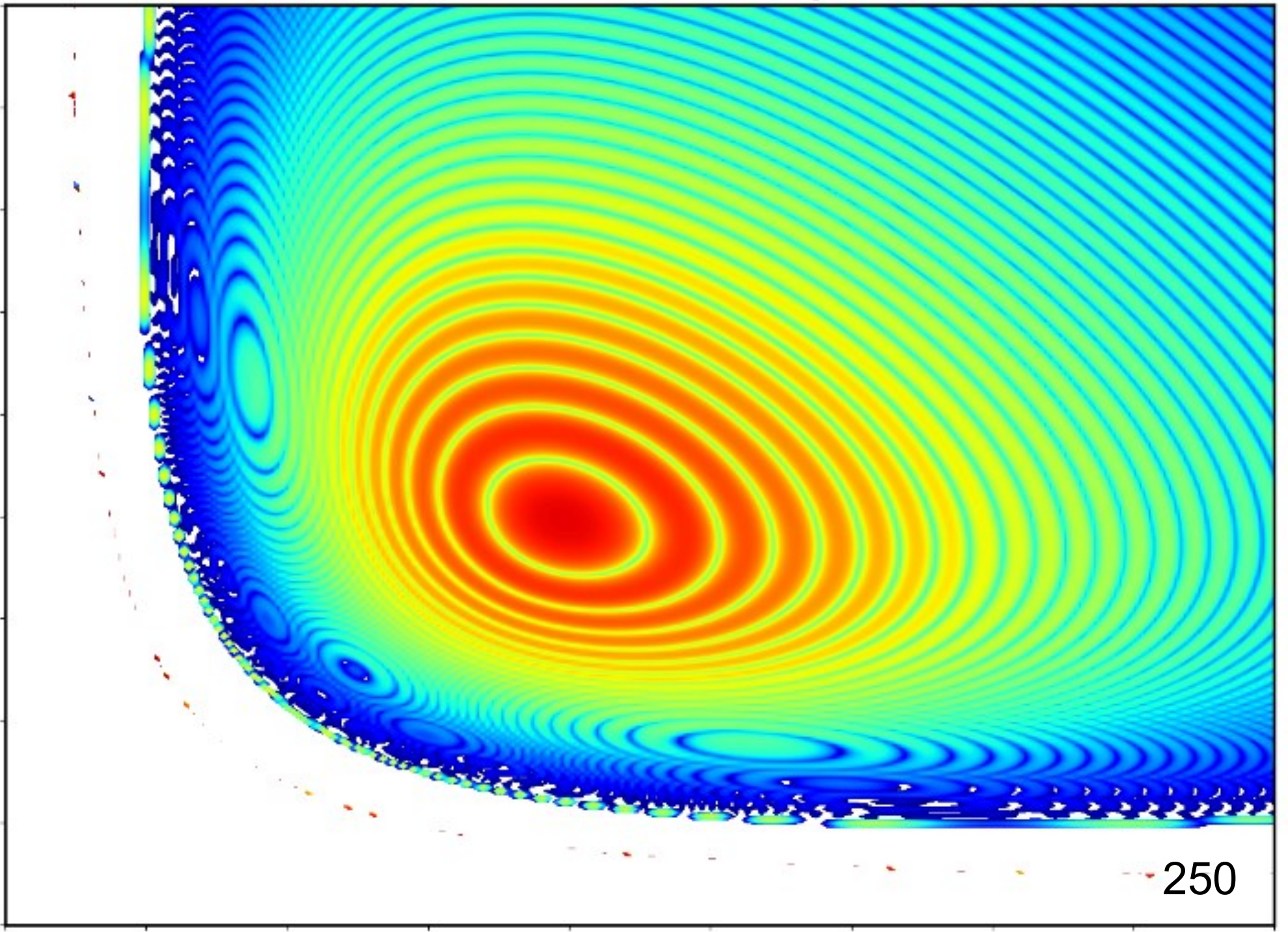} 
\caption{Dalitz type plot for trident pair production showing the differential rate as a function of final energies $E_2$ and $E_3$ for $a_0 = 10^{-4}$ and phase duration $\Delta \phi = 250$.} 
\label{fig:Kaempfer} 
\end{figure}

Both of the previous speakers displayed some first results of the final state momentum distributions (spectra), see e.g.\ Fig.~\ref{fig:Mackenroth}. Burkart K\"AMPFER (TU Dresden and HZDR) presented a wealth of such distributions (for the outgoing positron) as a function of boost invariant variables, namely transverse momentum, $p_T$ and rapidity $y$. In terms of these the positron 4-momentum is $p = (m_T \cosh y, \mathbf{p}_T, m_T \sinh y)$ with $m_T^2 \equiv p_T^2 + m^2$. He then studied the impact of pulse duration and intensity $a_0$ on the positron spectra: Reducing pulse length leads to broader distributions, i.e.\ an enlarged phase space, and positron yields increase with intensity. Another useful suggestion is to use Dalitz type plots (normally employed for 3-body decays into spin-0 particles). These represent differential production rates as a function of $s_{12} := (p_1 + p_2)^2$ and $s_{23} := (p_2 + p_3)^2$ with final momenta $p_1$, $p_2$ and $p_3$, see Fig.~\ref{fig:tridentRate}. A typical Dalitz plot is shown in Fig.~\ref{fig:Kaempfer}. 

For the nonlinear BW sub-process K\"AMPFER, coined the phrase `Ritus corner' for the parameter regime $a_0 \gg 1$, $\chi \ll 1$ leading to the asymptotic result (\ref{RITUS.CORNER}). He also pointed out that a comparison with the standard perturbative trident process ($a_0 = 0$) should be useful, and displayed a number of numerical spectra for this case. Finally, he highlighted the fact that a Dalitz plot is a well known tool to discover resonances, thus possibly new physics (more on this below).   

To summarise this subsection it seems fair to say that the trident process is coming under both analytical and, in particular, numerical control.

\subsection{Radiation reaction and consequences}

Stepan BULANOV (LBNL) reminded the audience of the `Bronstein cube' \cite{Bronstein:1933} of high-intensity physics with coordinate axes labelled by $1/c$, $\hbar$ and $a_0$ (suppressing the fourth axis for gravity, i.e.\ Newton's constant, $G$). When all three parameters are relevant, we enter the regime of \emph{high-intensity particle physics}, which remains poorly explored to this day. Facilities combining high-energy electrons with high-power lasers would take us straight into this regime. Introducing the classical radiation reaction (RR) parameter, $\epsilon_\mathrm{rad} := (2/3) \alpha b_0$, the speaker identified a regime where classical RR dominates energy loss due to quantum recoil. For a laser pulse containing $N$ cycles this holds as long as $N a_0^2 \epsilon_\mathrm{rad} = O(1)$ (substantial RR), while $b_0 \ll 1$ (low energy, no recoil). For typical optical frequencies this implies 
\begin{equation}
  a_0^2 \simeq \frac{3 \times 10^7}{\gamma N} \gg \frac{3 \times 10^2}{N} \;,
\end{equation}
the inequality implementing the low-energy requirement, which amounts to $\gamma \ll 10^5$. In terms of the parameter $\chi = a_0 b_0$, the RR regime is characterised by $\chi \simeq 5 \times 10^{-2} \sqrt{\gamma/N}$. Note that the DESY/XFEL beam has $\gamma \simeq 3.5 \times 10^4$, which takes us well into the quantum regime. 

BULANOV also recalled that deep in the quantum nonlinear regime ($a_0 \gg 1$, $\chi \gg 1$) one has to expect pair cascades or, more generally, QED particle showers. This may provide an ultimate upper limit for the laser intensity that can be reached on physical grounds. The pair creativity of the experimental set-up (even in the absence of an electron beam) may be further enhanced by sophisticated laser beam shaping and/or by utilising multiple colliding pulses. For a particular beam model, the number of positrons increases from $10^{-19}$ (exponential suppression) to $10^6$ upon increasing the number of colliding beams from 2 to 24 \cite{Bulanov:2010ei}.         

Marija VRANIC (IST Lisbon) reported on her (essentially) numerical findings regarding quantum RR, pair production and acceleration, giving the first presentation based upon the use of a particle-in-cell (PIC) code (OSIRIS). This approach is somewhat complementary to the usual scattering picture employed in particle collisions such as the trident process above. A scattering picture is only interested in the asymptotic (incoming and outgoing) particles and the distribution of their quantum numbers (momenta, spin, polarisations, etc.) and neglects any real time dynamics during the scattering process (essentially assuming that the scattering region is highly localised in space and time). On the other hand, codes such as PIC or others based on transport equations (Vlasov, Fokker--Planck, Boltzmann) explicitly study and simulate real time dynamics, typically adopting a semi-classical picture with particle trajectories augmented by probabilistic quantum `noise' and `branchings' mimicking quantum vertices. It is not straightforward to map one approach to the other and get consistent answers \cite{Blackburn:2018sfn}. 

\begin{figure}[ht]
\centering
\includegraphics[scale=0.7]{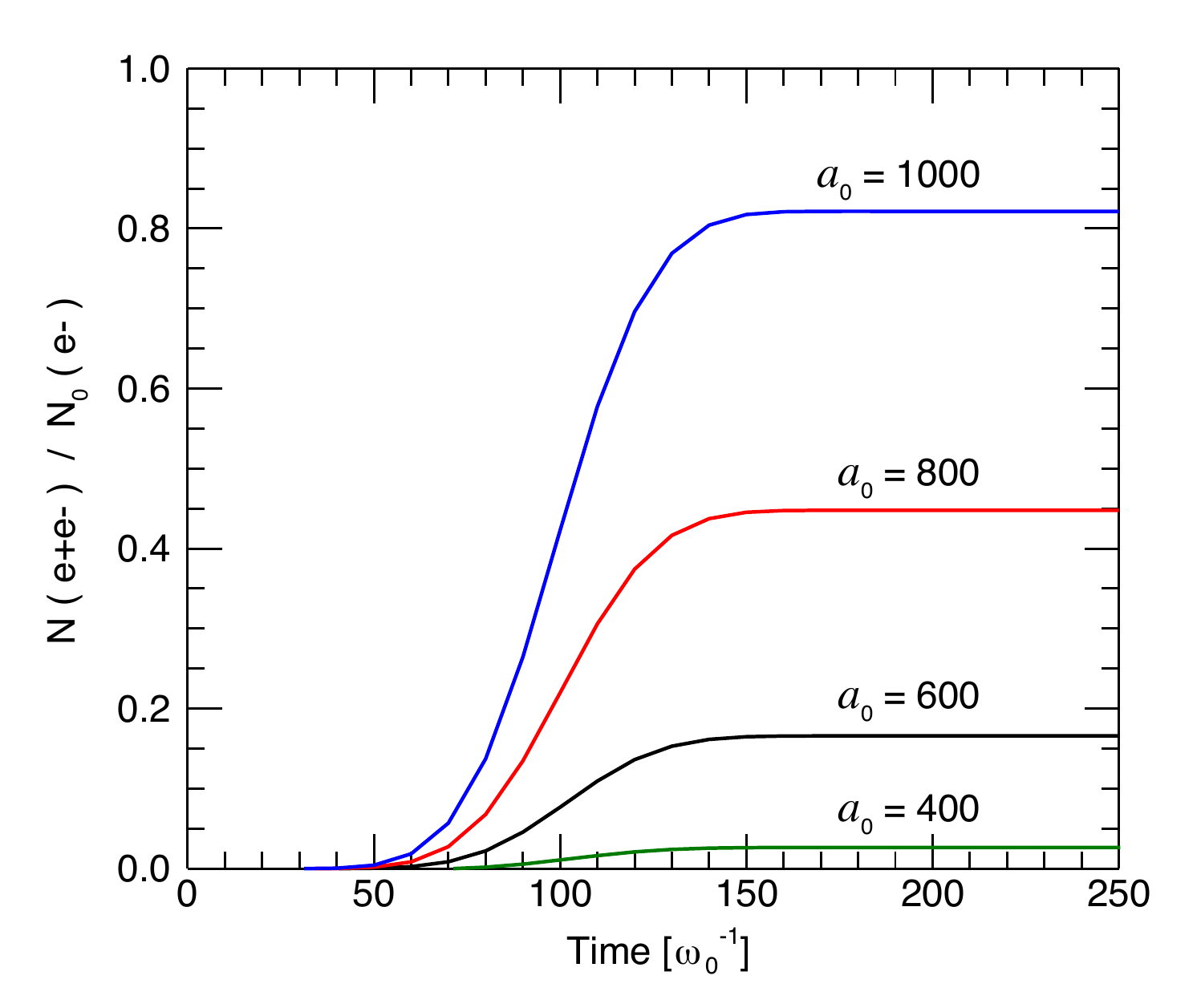} 
\caption{Ratio of pair number to initial electron number as a function of time for different values of $a_0$.} 
\label{fig:Vranic} 
\end{figure}

VRANIC discussed the influence of quantum RR on electron beam shape during collisions with a laser beam and identified diffusion and drift regimes at early and late times, respectively \cite{Vranic:2016myd}. In her second part she focussed on the generation of $e^+ e^-$ beams through electron laser scattering at 90 degrees \cite{Vranic:2018liw}. Due to laser defocussing, the generated pairs are accelerated in vacuum. The beam turns out to be quasi-neutral if the initial energy exceeds 2 GeV. For $a_0$ of the order of $10^3$, the number of pairs in the beam is comparable to the number of initial electrons, see Fig.~\ref{fig:Vranic}. 

Matteo TAMBURINI (MPIK, Heidelberg) presented the results of \cite{Benedetti:2017dmq} on generating ultra-bright gamma beams driven by instabilities of electron beams hitting a solid target, hence mediated by RR. Up to 60\% of the electron beam energy is converted into gamma rays via synchrotron emission in strong self-generated electromagnetic fields. Many possible instability modes are simultaneously excited and differ in their growth rates. These have been used to determine the evolution of both longitudinal and transverse electron beam number density, $n_b$, via PIC codes. Depending on the value of $n_b$, the self-generated magnetic fields can reach magnitudes in excess of 10 MG. For the largest value, $n_b = 6 \times 10^{20}$\,cm$^{-3}$, one achieves a brillance of a few times $10^{25}$ photons s$^{-1}$ mm$^{-2}$ mrad$^{-2}$ per 0.1 \% bandwidth and photon energies ranging from 200 keV up to several hundreds MeV. This should be compared with facilities (currently under construction) based on incoherent Compton back scattering which are expected to yield photon beams with energy up to 19.5 MeV and peak brilliance up to 10$^{23}$ photons s$^{-1}$ mm$^{-2}$ mrad$^{-2}$ per 0.1 \% bandwidth. 

\subsection{Beyond the Standard Model}

\begin{figure}[ht]
\centering
\includegraphics[scale=1.5]{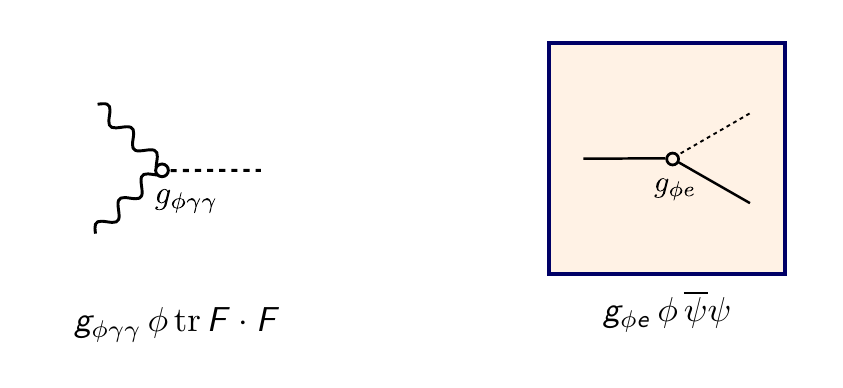} 
\caption{Left: Scalar coupling to photons. Right: Yukawa coupling of scalar to electrons. Interpreting the full lines as laser dressed electrons, the diagram represents the emission of scalars by an electron in a strong laser field.} 
\label{fig:King} 
\end{figure}

Ben KING (Plymouth) reported the findings of \cite{Dillon:2018ypt} and \cite{King:2018qbq} on massive scalar and pseudo-scalar production in electron--laser collisions. The particles produced could be axions or axion-like particles (ALPs) in the pseudo-scalar case or any weakly interacting slim particles (WISPs). The suggested experiments would probe for these and hence for physics beyond the Standard Model (SM), arising through couplings to the invariants $\mathcal{S}$ or $\mathcal{P}$ introduced in (\ref{SP}) above. In addition, the new particles could also interact with standard model leptons through Yukawa type interactions. The focus of the talk was on the latter possibility by considering  scalar emission by an electron in a strong laser field, see Fig.~\ref{fig:King} (right panel).   

The speaker discussed effects of the scalar mass on the production rates and compared the yields to that of pseudo-scalars. In the transitional regime, $a_0 \sim 1$, scalar production dominates over pseudo-scalar at low energy, i.e.\ when $b_0 < 1$. It turns out that coherence effects can drastically enhance low-energy scalar ALP production and thus one can put competitive lab-bounds on ALPs with masses above 1 eV.  

Selim VILLALBA-CHAVEZ (D\"usseldorf) gave an extended overview of their work on physics beyond the SM, in particular the relevance of laser-based experiments for probing hidden particles \cite{Villalba-Chavez:2016hht,Villalba-Chavez:2016hxw}. He pointed out that at energies near the scale specified by the electron mass, quantum vacuum fluctuations are dominated by the electron--positron fields. However, the source of quantum fluctuations inducing nonlinear self-interactions of the electromagnetic field is not restricted to virtual electrons and positrons. Quantum vacua characterised by much lower energy and electromagnetic field scales are related to very weakly interacting sub-eV particles, linked to theoretical frameworks beyond the SM. Indeed, there is a variety of SM extensions demanding the existence of mini-charged particles and ALPs. As a consequence, the corresponding vacuum polarisation effects might induce tiny distortions in the corresponding QED vacuum polarisation phenomenology. Hence, laser-based set-ups, designed to detect the hitherto unobserved QED vacuum birefringence, provide a genuine opportunity for probing these hypothetical degrees of freedom. In such set-ups the decay of probe photons into ALPs, WISPs etc.\ might induce a rotation of the polarisation plane of the associated small-amplitude electromagnetic wave, even though the corresponding probe photon energy is far below the threshold of electron--positron pair production. Consequently, the transmission probability through an analyser at right angle to the initial polarisation direction would not be determined solely by the QED ellipticity but also by the ellipticity and the rotation angle induced by these hypothetical particles. We argue that a slightly modified version of the X-ray polarimeter proposed by the HIBEF collaboration would allow measurement of both signals separately. A direct comparison between the outcomes associated with pulses with different envelope has indicated that the projected bounds resulting from the ellipticity are less sensitive to this dependence than the upper limits arising from a plausible rotation of the polarisation plane. Besides, the corresponding analysis suggests that a substantial part of the corresponding parameter space can be discarded, no matter what type of strong-field profile is utilised.

\subsection{Theoretical challenges}

The contribution by Holger GIES (FSU and Helmholtz Institute, Jena) on all-optical signatures from the quantum vacuum was originally meant for the synergies section (see below), but also fitted nicely under the headline of theoretical challenges.  He pointed out that intense lasers may also couple to virtual electrons (and positrons) via a 4-photon box diagram describing light-by-light scattering. At low energies, this can be quantitatively described through the Heisenberg--Euler effective action, $\Gamma_\mathrm{HE}$, and leads to effects such as vacuum birefringence, photon merging and splitting, 4-wave mixing etc. These can be most economically analysed in terms of a vacuum emission picture \cite{Karbstein:2014fva,Gies:2017ygp} which couples a classical current (the functional derivative of $\Gamma_\mathrm{HE}[A]$ with respect to the background $A$) to the emitted quantum field $\hat{a}$ of the photon. This results in a very simple formula for the signal photon count,
\begin{equation}
  dN (\mathbf{k}) = \frac{d^3 k}{(2\pi)^3} \left| \langle \gamma(\mathbf{k}) | \int d^4 x \, 
  j^\mu [A] \hat{a}_\mu |0\rangle \right|^2 \; ,
\end{equation}  
which can then be used for a detailed investigation of the collision of, say, Gaussian laser pulses, employing the locally constant field approximation. The latter assumes that the laser field does not substantially change across an electron Compton wavelength so that the Heisenberg--Euler low-energy approximation remains valid even for space-time dependent backgrounds. As a promising example, the collision of three pulses was re-analysed resulting in the prediction of 3.03 signal photons per shot \cite{Gies:2017ezf} employing a 3-beam configuration with energies of $25$ J and $2 \times 6.25$ J, respectively. 

Alexander FEDOTOV (MEPhI, Moscow) introduced the next great challenge of probing fully non-perturbative QED with electron--laser interactions. In doing so he gave an overview \cite{Fedotov:2016afw} of the Ritus--Narozhny conjecture that strong-field QED perturbation theory will break down when $\alpha \chi^{2/3} = O(1)$, i.e.\ at a magnitude of $\chi \simeq 1600$. When $\chi$ becomes that large, higher order loop diagrams will no longer be suppressed compared to the leading order, which suggests that all orders in strong-field QED perturbation theory have to be included -- a formidable challenge, to put it mildly. The fractional power-law behaviour in $\chi$ 
was nicely explained through a simple scaling argument based on the concept of loop size. Nevertheless, the physics involved remains unclear and seems to contradict the usual QED scaling of higher orders with $\alpha \ln \gamma$, which is logarithmic in energy. The second part of the presentation addressed the experimental feasibility of reaching asymptotic $\chi$ values which indeed seems possible in beam--beam collisions. The relevant parameters can be found in \cite{Yakimenko:2018kih} together with numerical results based on the simulation suite OSIRIS. An important conclusion was though that a viable theory is still missing in the (rather extreme) parameter range of interest. 

Mattias MARKLUND (Chalmers, Gothenburg) gave an outlook discussing the basic theoretical challenges to be met and came up with a list of four major issues or questions: (i)  How reliable is the locally constant crossed field approximation, which is central to almost all analytical work trying to include transverse focussing (including PIC codes)? In other words, what is the size of the errors that we are making using this approximation \cite{Ilderton:2018nws}? When precisely, e.g.\ for which amount of focussing, is it going to break down? (ii) Backreaction, in particular depletion \cite{Seipt:2016fyu}: How do we correctly take into account the partition of energy between classical and quantum degrees of freedom, and is it important? More generally, when and under which circumstances will the background field approximation become invalid? (iii) S-matrix vs.\ trajectories: When is it meaningful to talk about trajectories, and when is there a relation between the S-matrix and the classical equations of motion \cite{Blackburn:2018sfn}? Thus, in which parameter regime is a semi-classical approach justified? (iv) Many-body quantum systems: Are these computationally viable, and when will be a full quantum treatment be necessary? What are the methodological options (non-equilibrium quantum field theory, Schwinger--Keldysh formalism, real-time path integrals and Monte Carlo, Wigner functions, etc.). Carefully designed and fine-tuned experiments may be able to help answering at least some of these questions.


\section{Accelerators}
\label{sec:acc}
R. Assmann, F. Burkart

The LUXE workshop had a dedicated session on accelerator aspects with a total of five speakers.  During first part of 
this session the possible layout of the strong-field QED experiment LUXE at the EuXFEL was discussed, bringing 
together a new high-power laser and the multi-GeV EuXFEL electron beam. The EuXFEL layout and parameters 
were presented and the status of the LUXE design study was discussed. The second part of the session 
introduced the physics beyond colliders projects at CERN, laser-wakefield experiments and their possibilities to 
probe high-field QED. In summary, the presentations and discussions at this session set the scene for the possible 
LUXE experiment in Hamburg and put it into the international research landscape. 

Evgeny NEGODIN (DESY) described the European XFEL at DESY including the project history.  The general layout 
of the accelerator was shown. The EuXFEL consists of a 1.9\,km long superconducting accelerator to accelerate 
electrons up to 17.5\,GeV. There is the possibility to produce X-rays in the range of 0.25 to 25\,keV with three 
variable-gap undulators. The experimental area at Schenefeld can house six individual experiments.  The accelerator 
and the corresponding beam lines are located $6-38$\,m below the surface between the DESY main campus and 
Schenefeld. 

Negodin described the electron accelerator, showing that it can accelerate up to 2700 bunches/pulse with up to 1\,nC 
bunch charge. The different electron and photon beamlines were introduced and the corresponding experimental areas 
were shown. During user operation it is possible to extract single bunches from the bunch train with a fast kicker magnet 
and to send these bunches to a dedicated experiment. This possibility would be very beneficial for a strong-field QED 
experiment with low repetition rate.

Florian BURKART (DESY) introduced the present layout of LUXE.  A possible location for this strong-field QED experiment 
was found at the end of the electron beam line close to XSDU1.   Burkart discussed the kicker and septa magnet system 
that would extract one single bunch out of the bunch train.  This bunch will then be guided into an experimental hall via a 
focussing triplet magnet system and dipole magnets, that were described. Two experimental scenarios were presented: 
the electron beam directly interacts with the laser pulse; and the electron beam hits a foil, generates photons and these 
photons interact with the laser pulse.   Already existing magnets or magnet designs from the EuXFEL can be re-used. The 
layout and the achievable spot size at the interaction point are demanding but feasible. The presentation ended with an 
outlook towards the required decisions and further design and integration studies.  Burkart pointed out that the required 
time for the modification at the EuXFEL and its impact on EuXFEL operation are important and need to be evaluated with 
more detailed studies.

Mike LAMONT (CERN) introduced the Physics Beyond Colliders projects. This is an exploratory study aimed at exploiting 
the full scientific potential of CERN's accelerator complex and its scientific infrastructure through projects complementary 
to the LHC, HL-LHC and other possible future colliders.  There is a strong physics motivation with such projects to search 
for: light dark matter (LDM); portals to a hidden sector (HS) (dark photons, dark scalars); axion-like particles (ALP);  heavy 
neutral leptons (HNL);  and lepton flavour violating $\tau$ decays.  An already active (and continuously growing) set of 
experiments are ongoing at the intensity frontier, including projects at CERN (NA62, NA64, and SHiP), in Japan 
(BELLE-2) and in the US (LDMX, APEX, SeaQuest, HPS).

An update on the construction of SHiP (Search for Hidden Particles) using the 400 GeV proton beam from the SPS at 
CERN was given. The search for dark photons will be covered with the NA64 experiment (approved in March 2016). 
First design considerations for an eSPS to accelerate electrons up to 16 GeV and to slowly extract them to a test side 
on the Meyrin campus were shown. The CERN Axion Solar Telescope, where CERN contributed to magnets and R\&D 
is now operational. The proton-driven plasma acceleration proposal AWAKE++ has the potential to serve as a test area 
for fixed-target experiments to study dark photons, deep inelastic scattering and non-linear QED.

Gianluca SARRI (Queen's University Belfast) presented the results of the high-field QED experiments using the ASTRA 
Gemini laser at the Rutherford Appleton Laboratory. The experiment is split into low-intensity ($a_0 \sim 2$, 
$\gamma_0 \sim 1000$, $\chi \sim 0.01$) and high-intensity ($a_0 \sim 10$, 
$\gamma_0 \sim 4000$, $\chi \sim 0.2$) regimes. The experimental setup consists of a laser-driven electron accelerator, 
a laser--electron beam interaction point, electron spectrometer and an X-ray detector. The experiment had to solve issues 
from electron beam stability, pointing fluctuations and problems with the femtosecond timing synchronisation. The 
collision diagnostic was presented in detail. The comparison between theoretical models and measurement results at 
present lacks the knowledge of laser spectral phase and the longitudinal laser distribution.  For future experiments one 
would require higher laser intensity and higher electron energies, such as is foreseen in the EuPRAXIA design study.

Vitaly YAKIMENKO (SLAC) presented studies at SLAC with a high-energy facility for advanced accelerator R\&D (FACET).  
The 20\,GeV, 3\,nC accelerator was accessible since 2012 for users. The facility will be upgraded with the FACET-II 
program. Amongst others, this upgrade will enable studies that are preparing solutions for the energy frontier in future 
colliders, it enables experiments on electron beam emittance preservation and it enables the development of methods 
targeting brighter X-rays for photon science. The commissioning of the FACET-II RF gun started in 2018.  Yakimenko 
explained that the dense bunch of the electrons travelling in a plasma wakefield can break up into filaments, producing 
strong magnet fields. These fields will bend the electrons, leading finally to the emission of gamma-ray photons.   The 
presentation also discussed studies for the ILC interaction point. The results show that virtual particles would dominate 
the collisions in the ILC, leading to non-pertubative QED effects.


\section{High-power lasers}
\label{sec:lasers}
A.R. Maier

The workshop could attract high-profile speakers from the community of high-intensity laser development and applications, including 
representatives from ELI Beamlines, Lawrence Berkeley National Lab, Shanghai Institute of Optics and Fine Mechanics SIOM, and 
European XFEL, which presented an overview of the current state-of-the-art and discussed challenges to be addressed.

A general consensus of the discussion was that as the laser pulse properties at the interaction point (IP) directly determine the 
interaction rate, a precise characterization of the laser pulse at the IP is crucial for an accurate reconstruction of the Schwinger field.  
Laser pulse metrology on the scale of the required TW peak powers is, however, extremely challenging. In particular, no diagnostic 
devices exist that can handle the full peak power of the laser. Metrology of the laser pulse therefore relies on a carefully sampled 
and attenuated laser pulse with the requirement that the mechanism of sampling and attenuation over many orders of magnitude 
does not alter the pulse properties that are to be measured.

The state-of-the-art in high-intensity laser pulse characterization is just at the edge of what is required to appropriately characterize 
the pulse. From a laser point of view, the challenge of the LUXE experiment will not be on providing sufficient laser power on target, 
but to characterize the laser pulse with sufficient precision to not limit the data analysis accuracy.  This issue is, however, a general 
problem, which is shared by a broad field of applications of high-intensity lasers, including high-energy 
density user experiments at free-electron lasers, or plasma wakefield acceleration. There are strong efforts by the community to address 
the issue of laser pulse characterization, and, although challenging, the metrology of the high-power laser pulse for LUXE should be a 
manageable problem.

In addition to the discussion of laser pulse characterization, the speakers presented a broad overview of the current high-power laser 
activities, the available technologies and state-of-the-art performance.

Georg KORN (ELI Beamlines) provided a general overview of the available laser architectures and technologies that enable ultra-short 
pulse, high peak-power laser systems on the 1\,PW to 100\,PW scale and discussed specific challenges that are linked to the generation 
of such extreme pulses. While single-PW systems are already available, several facilities providing lasers on the 10\,PW scale and beyond 
are currently under commissioning. Concepts do exist that could eventually support not only peak powers on the TW and PW scale, but 
also average powers on a MW scale, enabling laser systems with very high repetition rates. Concepts for exawatt scale lasers are currently 
being discussed by the community.  An example of the currently most advanced laser systems in operation is the HAPLS/L3 laser available 
at the ELI Beamlines user facility close to Prague, which provides 1\,PW pulses (30\,J at 30\,fs) with a repetition rate of 10\,Hz. This 
cutting-edge laser will act as a workhorse for high-intensity laser experiments at ELI Beamlines.

Stepan BULANOV and Wim Leemans (both LBNL) presented the current status of the BELLA 1\,PW laser facility at Berkeley.  BELLA 
has demonstrated many pioneering results in the field of laser--plasma acceleration, including the first generation of 4\,GeV electron 
beams in a single plasma stage. Simulations show, that BELLA should be able to demonstrate a 10\,GeV electron beam from a single 
plasma stage in the near future. These challenging experiments require precise modelling of the plasma target (CFD and MHD 
codes) and guiding of the laser pulse over many Rayleigh lengths, using a laser-heated plasma channel.

Toma TONCIAN (Dresden) presented the status of the HiBEF laser, which is currently being commissioned at the high energy density station 
of European XFEL. The HiBEF user consortium aims to provide this highly sophisticated instrumentation to the whole HED community.  
Experiments under extreme pressure, field, plasma and temperature conditions will be explored, using a 300\,TW (7.5\,J, 25\,fs) laser and 
a 100\,J, 20\,ns laser. Integration of such laser systems into the user operation of European XFEL are technologically challenging and the 
expertise of HiBEF may prove very useful for the later integration and commissioning of a LUXE laser into the European XFEL infrastructure.

Liangliang JI (Shanghai) presented experiments and plans for the Station of Extreme Light (SEL) using a 100\,PW laser system at Shanghai.  
SEL is located in close vicinity to the Shanghai XFEL and aims to provide 1.5\,kJ pulses with 15\,fs duration at $10^{23}$\,W/cm$^2$ intensity, 
with the goal of enabling high energy density physics and strong-field QED experiments. This includes photon--photon collision studies and 
detection of vacuum birefringence. Colliding the laser with an electron beam enables radiation-reaction studies and generation of 
high-brightness gamma beams.

Georg KORN and Sergei BULANOV (both ELI Beamlines) presented an overview of the high-field experiment activities at ELI Beamlines.  
Those experiments will be operated by the HAPLS/L3 laser (see above) and the L4 ATON laser, providing 10\,PW pulses (1.5\,kJ, 150\,fs) 
at 1 shot per minute repetition rate. It is planned to generate gamma-ray flashes from the interaction of the lasers with a solid target, and to 
prove the vacuum polarization with Cherenkov-Compton radiation processes. The physics and scaling laws of reaching and going beyond 
the Schwinger field limits were presented.

Andreas MAIER (Hamburg University) presented latest results from the LUX laser plasma accelerator, which is operated with a relatively 
modest laser power of 100\,TW. The focus of the facility is to generate electron beams from a plasma accelerator with high availability and 
reproducibility, which, however, relies crucially on a laser system operated with high availability and reproducible laser pulses.  A significant 
improvement in the performance of the LUX laser-plasma accelerator could be achieved by the integration of the drive laser system into the 
accelerator infrastructure and, in particular, the accelerator controls and data acquisition system. The lessons learned from this facility could 
be beneficial for the design and operation of a LUXE laser system.


\section{Experimental set-up and detectors}
\label{sec:det}
T. Koffas, D. Reis.

In this section, a brief investigation of the available detector technologies for a strong-field QED experiment using a high-power laser and a multi-GeV electron beam Linac, are investigated.
A brief overview of the laser and electron beam parameters and how they translate to the parameters of interest $\chi=\Upsilon$ and $\xi=a_0$, is presented. This is followed by a description
of the expected interaction products and their rates. Based on this information then, the available detector technologies are investigated and evaluated.

\subsection{Initial beams}

Initial electron bunches are assumed to be like those found at EU.XFEL (LUXE) and SLAC (LCLS/FACET-II) and approximate values of the relevant parameters are:
\begin{itemize}
\item an electron energy, $5 < E_e< 20$\,GeV;
\item a bunch charge of $N_e = 10^8 - 10^{10}$\,electrons;
\item a bunch length of $\sigma_z \sim$ few $- ~100\,\mu$m;
\item the repetition rate will be dictated by the laser frequency;
\item a beam size of $\sigma_{x,y} \sim$ few $- ~50\,\mu$m.
\end{itemize}

The high-power laser is assumed to have the following properties:
\begin{itemize}
\item a 100\,TW [1\,PW] laser system, e.g.\ 2.5\,J [25\,J] in 25\,fs;
\item a typical focal area of ($10\,\mu$m)$^2$, implying an intensity of $10^{20}$\,W\,cm$^{-2}$ [$10^{21}$\,W\,cm$^{-2}$];
\item a Ti:sapphire laser system (800\,nm central wavelength, i.e.\ typical photon energy: 1.55\,eV);
         this implies a reduced vector potential of $a_0 = 5 [a_0 = 15]$ and a quantum parameter of $\chi = \Upsilon \approx 0.6$  $[\chi=\Upsilon \approx 1.8]$ for head-on collisions with a 10\,GeV electron/photon\footnote{The properties of SFQED processes in a laser pulse with photon energy $\hbar\omega_L$ and electric field strength $E$ are mainly determined by two gauge and Lorentz invariant parameters,
\begin{gather*}
	\chi = \Upsilon = \frac{2\epsilon}{mc^2} \frac{E}{E_{\text{cr}}}  \approx 0.5741 \, \frac{\epsilon}{{\rm GeV}}\, \sqrt{ \frac{I}{10^{22}\,{\rm W / cm^{2}}}},
\quad
	\xi = a_0  = \frac{eE}{mc\omega_L} \approx 0.7495 \, \frac{{\rm eV}}{\hbar\omega_L} \sqrt{ \frac{I}{10^{18}\,{\rm W / cm^2}}}.
\end{gather*}
[$e>0$ denotes the elementary charge, $m$ the electron/positron mass and $E_{\text{cr}} = m^2c^3/(e\hbar) \approx 1.3 \times 10^{18}\,{\rm V / m}$ the QED critical field]. The expression given for $\chi$ holds only for a head-on collision of an (ultra-relativistic) electron/positron or photon with energy $\epsilon$.}
\item the repetition rate will be in the range $0.1 - 10$\,Hz;
\item the laser is linearly polarized (full control of the polarization, i.e. linear over elliptical to circular possible), thus there is the possibility of producing polarized photons and electrons and positrons;
\item near backscattering with electron and/or gamma beam;
\item some level of frequency conversion and longitudinal pulse shaping will be available at reduced intensities.
\end{itemize}

\begin{figure}[ht]
\centering
\includegraphics[scale=0.5,trim={0cm 7cm 0cm 3.5cm},clip]{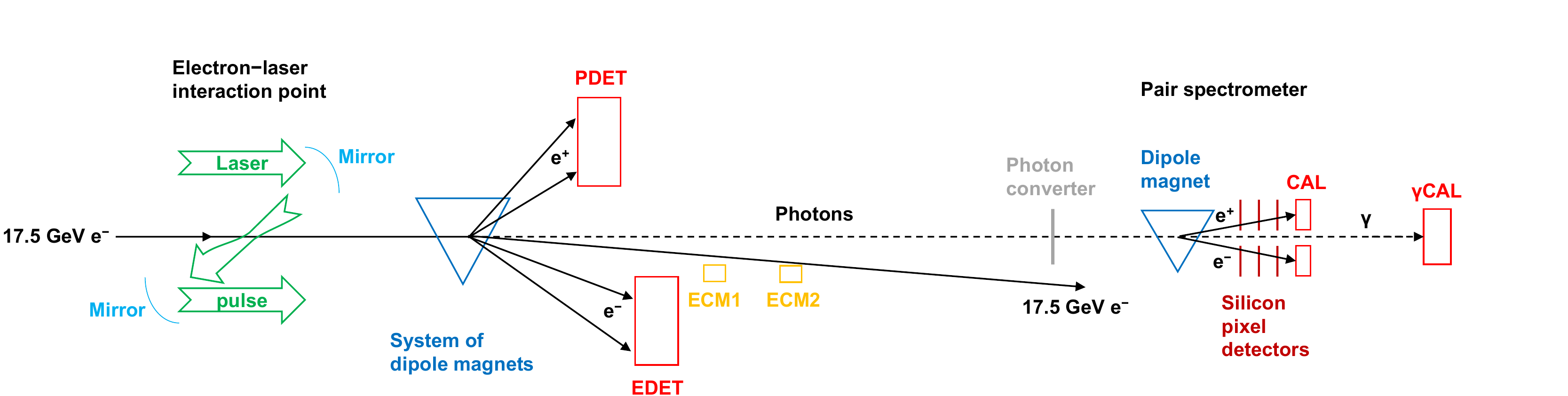}
\caption{Experimental arrangement of a future strong-field QED experiment where a high-power laser beam is brought into collision with an electron beam produced at a linac.} 
\label{fig:expt-layout} 
\end{figure}

A potential experimental arrangement could look like the one shown in Fig.~\ref{fig:expt-layout}.

\subsection{Final-state particle properties}

\underline{\textbf{Photons:}}
With the significant increase in laser intensity, we expect a copious production of scattered photons:
\begin{itemize}
\item up to $\sim 10^{11}$ photons, i.e.\ up to 10 photons per incident electron;
\item the transverse size of the photon bunch at the interaction point will be about $10\,\mu$m, i.e.\ $<$\,mm for $\mathcal{O}(10$\,m) downstream (can change with different emittance of electron beam);
\item typical energy will be $1 - 10$\,GeV.
\end{itemize}
The main objective here will be to measure the energy spectrum of the produced photons and compare this with the spectrum measured for the Compton-scattered electrons. The two spectra should be fully complementary
and provide a very clear measurement of the interaction's available centre-of-mass energy. This in turn should allow for a precise determination of the electric field strength at the interaction point and hence determine
whether the perturbative on non-perturbative regime has actually been reached.

\noindent \underline{\textbf{Electrons:}}
The final-state electrons consist of three different populations: 
a) electrons which have never or at most radiated soft photons (nearly unchanged from the initial electrons); 
b) electrons which have radiated a hard photon (energy losses up to 99\% are feasible); 
c) electrons which are produced via photon decay into electron--positron pairs (total energy $\geq 1$\,GeV, due to threshold). 
Furthermore, the electrons differ in transverse momentum, which is induced by the laser. Due to the Lawson--Woodward theorem the final state of population a) is nearly unchanged. 
Electrons from population b) and c) exhibit a transverse momentum of $\sim 5$\,MeV along the laser polarization direction and $\sim 0.5$\,MeV along the orthogonal direction 
(for circular polarization there is no preferred direction and the transverse momentum is of order 5\,MeV everywhere). We expect in $e/ \gamma$ collisions:
\begin{itemize}
\item $10^{10}$ electrons, i.e.\ one electron per incident electron;
\item the electrons will be separated from the positrons and will also have some spread due to a magnetic spectrometer;
\item typical energy will be $1 - 10$\,GeV.
\end{itemize}
By measuring the electron energy spectrum, one can cross-check the photon signal (and vice-versa). Obviously one would want to distinguish the electrons in pair production from those in Compton scattering.
Ideally one would like to measure both the energy and the momentum of the final state electrons. Also ideally one would like to correlate pair electrons to their positron partners, either via timing coincidence 
techniques or perhaps more elegantly via associating them to a common production vertex originating at the interaction point. The latter will again provide a very precise determination of the available
centre-of-mass energy for the inelastic light-by-light scattering and hence determine the perturbative (or not) nature of the underlying interaction.

\noindent \underline{\textbf{Positrons:}}
The number of electron--positron pairs depends strongly on the input parameters such as the laser pulse length and Lorentz invariants that characterize the strong-field interaction. 
At low fields, an electron--positron pair will be produced per bunch crossing. At higher fields, up to 0.1 electron--positron pairs are expected per incident $e/ \gamma$. We expect:
\begin{itemize}
\item $1 - 10^9$ positrons, i.e.\ up to 0.1 electron-positron pair per incident electron;
\item the positrons will be separated from the electrons and will also have some spread due a magnetic spectrometer;
\item typical energy will be $1 - 10$\,GeV.
\end{itemize}
One could then envision going from a scenario producing low numbers of pairs where we would want to measure both the $e^+$ and $e^-$ separately and correlate them, 
to one where much higher pair-production rates are achieved and such correlation measurements are no longer feasible.

\subsection{Semiconductor detectors for tracking/vertexing}
Reconstructing the trajectory and hence the momentum of the final state electrons/positrons, their production vertex and potentially correlate electrons to their positron partners from a pair, are highly desirable measurement capabilities.
To achieve this one would require high spatial resolution (defined by pixel size), excellent signal-to-noise performance even after extended exposure to high levels of radiation, 
low mass to minimize loss of trajectory information due to multiple scattering and in the case of vertex reconstruction a layered detector scheme where the first layer is as close to the interaction point as possible. 
Semiconductor trackers as presented by Chris KENNEY (SLAC) can provide these capabilities but their applicability may be severely limited by the number of produced charged particles. 

Hybrid designs are the dominant paradigm at the LHC, synchrotron and FEL facilities and characterized by having the sensor and front-end readout circuitry on separate silicon substrates. 
A typical example is the ePix hybrid detector~\cite{epix1,epix2} already employed at SLAC's LCLS facility.
Active CCDs, potentially cooled to lower temperatures, offer an important alternative due to their lower mass achieved by placing the front-end circuitry outside the active area. 
A CCD vertex detector was used by SLD~\cite{ccd-sld} at SLAC and was also part of the E-144 experiment's magnetic spectrometer setup. In recent years CMOS monolithic active image 
sensors~\cite{maps} are being actively pursued by experiments at the LHC or for future particle colliders. They combine the advantages of combining sensor and circuitry in the same substrate 
with low noise and inherent radiation hard performance. At high numbers of final-state charged particles, they may provide the only viable option for a tracking detector.

The applicability of semiconductor detectors for track/vertex reconstruction processes is mostly determined by the number of the produced electrons/positrons at the interaction point. One can define three such regimes:
\begin{itemize}
\item Low intensity, i.e.\ fewer than $10^5$ electrons per event: One can track every electron/positron produced. Existing detector technologies in the form of CCDs, monolithic or hybrid sensors can be used
         and tracking/vertexing configurations providing patterns by correlating hits across planes can be used. 
\item Moderate intensity, i.e.\ fewer than $10^8$ electrons per event: One can use an autoranging hybrid detector, e.g.\ ePix10K, AGIPD~\cite{agipd}. It could potentially require
         sensor redesign to optimize well depth versus noise. A tracking arrangement should still be able to resolve at the single particle level assuming $1-10$ megapixels per tracker plane.
\item High intensity, i.e.\ above $\sim 10^8$ electrons per event: The ability to reconstruct single charged-particle tracks is lost. In addition radiation damage and effects on long-term signal-to-noise
         levels and gain becomes worrisome. Monolithic CMOS tracking detectors but with low-efficiency transducer/sensor may be the only viable option. Even recording an integrated image could
         be challenging
\end{itemize}

\subsection{Calorimeters and Monitoring Detectors}
\noindent \underline{\textbf{Calorimeters:}} To improve the precision of the strong-field QED measurements and in particular to achieve full reconstruction of the available centre-of-mass energy at the interaction point,
a careful measurement of the produced electron/positron as well as photon energies is required. At the E-144 experiment at SLAC, this was achieved by using Si--W sampling calorimeters. Such a technology is able to 
satisfy the main requirements on the calorimeter detector performance, namely, good energy resolution, high dynamic range and modest impact resolution which in turn combined with the tracking/spectrometer information,
could allow for an estimation of the number of final state particles. The high dynamic range is necessary to accommodate for the variable numbers of produced final-state particles ranging from a few to as many as $10^{10}$
depending on the experimental conditions at the interaction point. The same technology could be used for the detection of both electrons and photons, to be characterised by a relatively small Moli\'{e}re radius and hence
narrow transverse electromagnetic shower and linear energy response in the $1-10$\,GeV energy range of interest. Timing information is an additional desirable functionality that could provide further handles to separate
signal from background and to correlate the simultaneously detected electrons/photons and thus perform a fully reconstructed final-state particle energy measurement. 

Semiconductor-based calorimeter detectors appear to provide the most promising avenue towards a full scale future strong-field QED calorimeter also presented by Chris KENNEY. One example are auto-ranging semiconductor arrays developed for
XFEL experiments such as AGIPD, LPD, ePix10K. Typical electron noise can be kept low even at high gains by having each pixel to automatically select appropriate gain levels on a pulse-by-pulse basis. Other
examples reported by Sasha BORYSOV (Tel Aviv) include thin silicon sensors sandwiched within submillimetre gaps between tungsten absorbers developed for the ILC such as the LumiCal prototype that has demonstrated good linear response in
the $1-5$\,GeV energy range and less than a millimetre impact position resolution~\cite{lumical}.

\noindent \underline{\textbf{Cherenkov Monitors:}} In any future strong-field QED experiment, it will be extremely important to monitor the quality of the interaction region and in particular the ability of the incoming beam electrons
to actually traverse the strong-field region created by the focused high-power laser beam and indeed maintain stable strong-field interaction conditions throughout a typical data collection campaign. An elegant way of achieving this 
is through the use of Cherenkov monitors located downstream the interaction point dedicated to the detection of electrons within a narrow energy range indicative of the order of the non-linear Compton scattering which produced them.
These monitors can also be used during the initial stages of the experiment to provide the necessary diagnostics in order to optimize the spatial and temporal overlap of the electron and laser beams.

\begin{figure}[ht]
\centering
\includegraphics[scale=0.5]{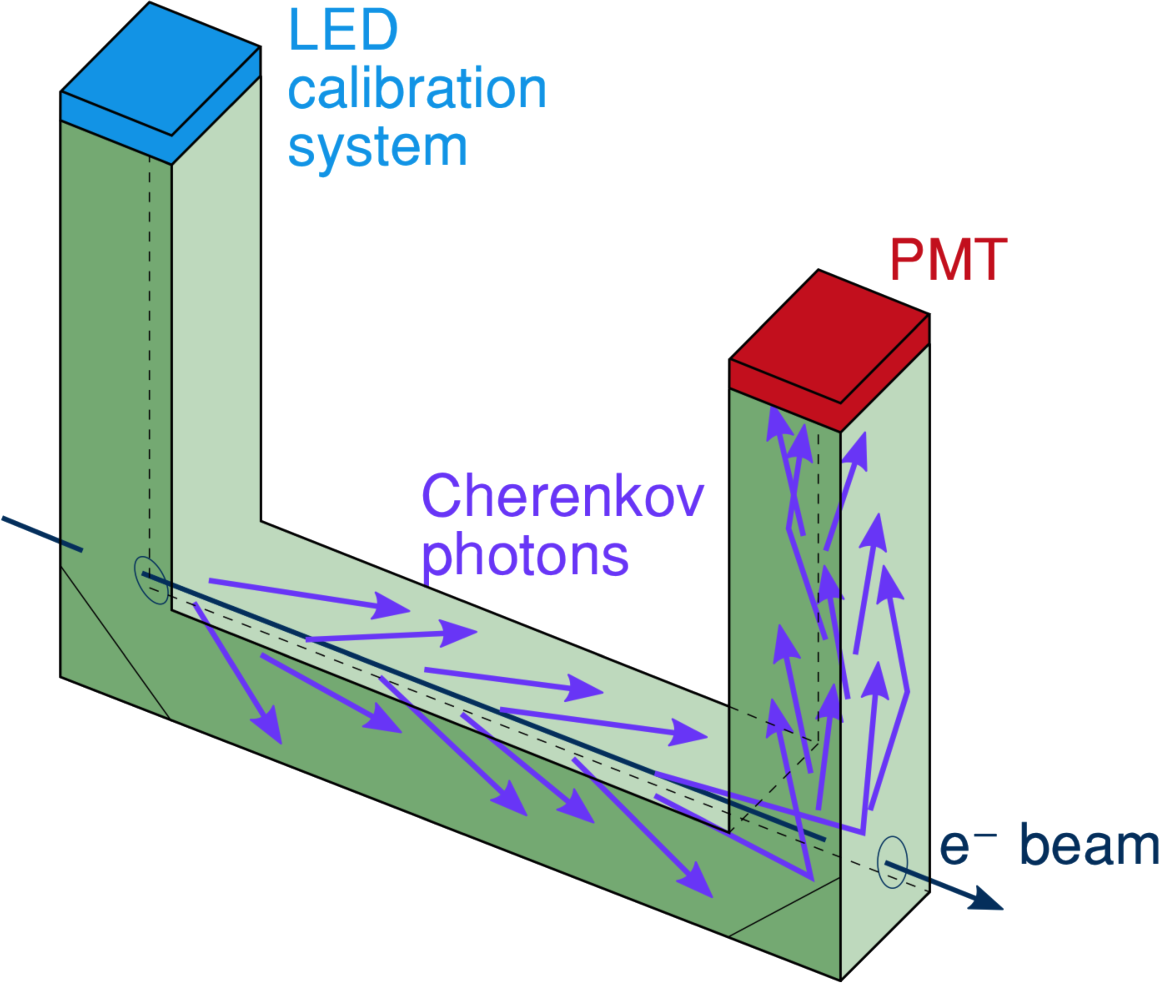}
\caption{Schematic of a typical Cherenkov detector arrangement to be used in a strong-field QED experiment at a linac.} 
\label{fig:cherenkov} 
\end{figure}%
A typical schematic of a Cherenkov detector is shown in Fig.~\ref{fig:cherenkov} as reported by Jenny LIST (DESY). Typical Cherenkov media (gas, quartz) are intrinsically radiation hard and could easily handle fluxes of $10^7$ electrons/positrons per second.
By introducing a Cherenkov threshold, low energy backgrounds can be suppressed. Finally, sensitive parts, such as photodetectors, can be placed outside of the beam plane.
By combining Cherenkov detectors with a magnetic chicane arrangement, they could be extremely useful to measure the relative polarization of the interacting electron and laser beams, of particular interest when, for example,  
electron-spin effects on theoretical predictions are studied. This can be achieved by measuring the asymmetry of the rate of the Compton-scattered electrons within a predetermined energy range, a quantity directly dependent 
on the relative beam polarization. Indeed, strong-field QED experiments could benefit from the on-going R\&D on Cherenkov detectors to perform polarimetry~\cite{cherenkov1,cherenkov2,cherenkov3} on a future linear collider.


\section{Synergetic projects}
\label{sec:syn}
M. Altarelli.

The achievement of very large electromagnetic fields, opening the way to non-perturbative strong-field QED phenomena is a topic of very high current interest, that attracts many groups worldwide with a variety of backgrounds: from high-power lasers and quantum optics, to accelerator physics, to nuclear and sub-nuclear physics, to plasma physics. The purpose of this session is to attempt an overview of proposed or recently carried out experiments, using approaches beyond the mainstream use of  high power lasers and/or relativistic electron beams.

The first two speakers reported on the search for light-by-light scattering, a quantum mechanical process that is forbidden in the classical theory of electromagnetism (linearity of the Maxwell equations), but is predicted to take place in QED.

Mateusz DYNDAL (DESY) reported on the measurements of light-by-light scattering at LHC \cite{ATLAS}. The collaboration analysed data from $Pb - Pb$ ion collisions at a centre-of-mass energy per nucleon pair of $5.02$\,TeV by the ATLAS detector. A total of 13 candidate events were observed with an expected background of $2.6 \pm 0.7$\,events. After background subtraction and analysis corrections, the fiducial cross section of the process $Pb + Pb (\gamma \gamma) \rightarrow Pb^{(*)}+ Pb^{(*)}\gamma \gamma $ , in the selected experimental conditions (photon transverse energy $E_{T} > 3$\,GeV, diphoton invariant mass greater than $6$\,GeV, etc., see Ref. \cite{ATLAS}) was measured to be $70 \pm 20\,{\rm (stat.)} \pm 17\,{\rm (syst.)}$\,nb, to be compared with Standard Model predictions of $45 \pm 9$\,nb~\cite{Enterria} and $49 \pm 10$\,nb~\cite{Gawenda}.

Toshiaki INADA (U. of Tokyo) reported on the build-up of light-by-light scattering experiments at the Japanese SACLA free-electron laser (FEL) facility. A configuration based on the collision of two X-ray beams produced by the FEL is not very practical, due to limitations of the efficiency of X-ray optics, and a more promising approach involves the collision of a FEL beam with a high power optical laser beam (a PW-class laser is envisaged, focussed to a $\simeq 1-3\,\mu$m spot). Assuming a similar focus for a $10$\,keV FEL beam, the expected count rate is $\simeq 10^6$ events per day. This requires however an optimized space and time overlap between the optical and the X-ray pulses; as part of the development of the set-up for the experiment, the apparatus to implement this overlap was tested using a 2.5\,TW laser, and proved pulse superposition to the $\mu$m level in space and $\simeq 100$\,fs in time. The observed counts matched the predictions for the corresponding FEL and laser intensities. 

In a similar vein, using the high power X-ray beams from FELs to search for high-field QED phenomena, Matthias FUCHS (U. of Nebraska-Lincoln) reported on non-linear Compton scattering experiments on Be targets, carried out at SLAC \cite{Fuchs}. In these experiments, two hard X-ray photons with energies around $9 keV$ scattered into a single higher-energy photon, red-shifted from the second harmonic, the missing energy transferred to electrons from the Be target. The maximum intensity was $\simeq 4 \times 10^{20}$\,W/cm$^{2}$,  corresponding to a peak electric field of $\simeq 5 \times 10^{11}$\,V/cm. The strong X-ray fields were produced in the $ \simeq 100$\,nm focus of the linearly polarized Linac Coherent Light Source XFEL, with pulse energy and duration $ \simeq 1.5$\,mJ, $50$\,fs, using the Coherent X-ray Imaging (CXI) instrument. The number of scattered  photons varies quadratically with the FEL intensity, as expected for a second-order nonlinear process, and is well above the measured background. A puzzle connected to this experiment is that the photon spectrum shows an anomalously large redshift in the nonlinearly generated radiation compared to the free-electron theory. Observations are incompatible with the customary impulse approximation, that should be very accurate in this case, where the X-ray photon energies are approximately two orders of magnitude above the Be 1s electrons binding energy.

David REIS (SLAC) presented another scheme, developed in collaboration with C. Pellegrini, to exploit the high power of X-ray FEL beams: colliding a powerful X-ray pulse from the LCLS with an electron bunch accelerated to $\simeq 6.5$\,GeV. To enhance the intensity of the FEL X-ray pulse, they proposed a seeding scheme with the so-called "double bunch" technique~\cite{Emma}. A first bunch generates a "seed" FEL pulse in a first undulator, that is monochromatized and delayed by crystal reflections, then superposed to a second bunch (trailing the first one by $1$\,ns) so as to generate a TW peak power pulse. This $8$\,keV X-ray pulse is then reflected and focused to a $\simeq 15$\,nm spot, with a peak power density attaining $\simeq 4 \times 10^{23}$\,W/cm$^{2}$, where it collides with a third electron bunch. In the rest frame of this latest bunch the electric field exceeds the Schwinger limit by a large factor, although the normalized intensity is reduced by the larger frequency (shorter wavelength) of X-rays with respect to an IR or optical laser. The X-ray optics achieving such a small focal spot at this high incoming power is very demanding, but the proponents envisage a high scientific pay-off (including observation of Unruh radiation in the lab) to motivate the necessary efforts.

Ulrik I. UGGERH\O J (Aarhus U.) reported on a completely different approach to reach the regime of strong field QED. He described experiments on channelling of very high energy electrons or positrons ($50$ to $180$\,GeV) from the CERN SPS into aligned crystals (e.g.\ Si or Ge). A simple analysis of the charge density in the channels between atomic rows in the crystals shows that, at distances of a few tenths of \AA \ from the atoms, a field of order $10^{10} - 10^{11}$\,V/cm exists; with the Lorentz factor $\gamma$ of order $10^5$ corresponding to the high energy of the electrons/positrons, the field seen in their rest frame is easily comparable or exceeding the Schwinger critical field. Transverse acceleration of the beams by such fields produces high energy $\gamma$'s by synchrotron radiation \cite{Ugge1}; this is indeed a rare case in which the {\it quantum limit} of synchrotron radiation emission is experimentally accessible in the lab (usually synchrotron radiation from storage rings is in the {\it classical limit}, in which the momentum of the emitted photons is very small compared to the momentum of the orbiting electrons \cite{Schwinger}). Apart from being a test of strong-field quantum electrodynamics, these experimental results are also relevant for the design of future linear colliders where bremsstrahlung, a closely related process, may limit the achievable luminosity. Another important experiment \cite{Ugge2} measured radiation emission spectra from ultrarelativistic positrons in silicon, in a regime where quantum radiation reaction effects dominate the positron's dynamics, i.e.\ where the dynamics is strongly influenced not only by the external electromagnetic fields, but also by the radiation field generated by the charges themselves, and where each photon emission may significantly reduce the energy of the particle.

Valery TELNOV (Novosibirsk) discussed ideas for $\gamma \gamma$ colliders as an add-on for linear electron/positron accelerators, a concept which has been around for many decades, but never found practical application; this is to a large extent explained by the postponement of implementation of large scale linear colliders. In view of this situation, Telnov investigated the possibility of a parasitic use of the European XFEL $17.5$\,GeV superconducting linac. A scheme was presented in which spent electron bunches are extracted and led through semi-circular arcs, where $e^{-} \rightarrow \gamma$ conversion is implemented by inverse Compton scattering on visible photons from a TW laser. Colliding photons would cover the $W_{\gamma \gamma} \leq 12$\,GeV region. Studies of precision spectroscopy of $c\overline{c}$ (charmonium) and $b\overline{b}$ (bottomonium) are among the proposed scientific payoff areas for the facility.

In addition to these experimental reports, Thomas GRISMAYER (U. of Lisbon) presented QED particle-in-cell (PIC) simulations of laser absorption via cascades of hard photon emissions and pair generations in the focus of counterpropagating laser pulses. The lasers are in the $\simeq 10$\,PW peak power range that should be realistically available in the near future, and the assumed focal spots are in the few $\mu m$ range. Simulations were performed via the OSIRIS 3.0 package and explored the intensity and polarization dependence of the growth rate~\cite{Gris1,Gris2}. In appropriate limits, comparison with analytical models is possible and provides a benchmark for the simulations. The laser energy is mainly absorbed due to hard photon emission via nonlinear Compton scattering. The results show that relativistic pair plasmas and efficient conversion from laser photons to $\gamma$ rays can be observed with the typical intensities planned to operate on future laser facilities such as ELI or Vulcan.

In summary, the session showed the impressive convergence of efforts from accelerator research, free-electron lasers, high-energy physics facilities, high-power laser developments, plasma physics and theory and simulation towards the goal of access to uncharted areas of  strong field QED.


\section{Summary and conclusions}
\label{sec:summary}

This workshop demonstrated that strong field QED is an area of physics with much scientific interest.  It is a region relatively unexplored 
in the laboratory and so should be measured, thereby challenging our current understanding and searching for new physics.  However, 
strong fields do exist in many natural systems or phenomena and so measuring them in the laboratory can lead do a deeper understanding 
of neutron stars, atomic physics, colliding bunches of particles, etc.  Various experiments are proposed worldwide and the next few years will 
be an exciting time for this field in which the region up to and beyond the Schwinger critical field will be investigated. 

\pagebreak[4]
\

\end{document}